\documentclass[pra,twocolumn,groupedaddress,showpacs,floatfix]{revtex4}

\usepackage{graphics}
\usepackage{graphicx}
\usepackage{bm}
\usepackage{amsmath}
\usepackage{amsfonts}
\usepackage{amssymb}
\usepackage{latexsym}
\usepackage{color}
\usepackage{epstopdf}

\newcommand{\beq}{\begin{eqnarray}}
\newcommand{\eeq}{\end{eqnarray}}
\newcommand{\bal}{\begin{align}}
\newcommand{\eal}{\end{align}}
\newcommand{\beqn}{\begin{dmath*}}
\newcommand{\eeqn}{\end{dmath*}}

\begin{document}

\title{Mollow triplet for cavity-mediated laser cooling}
\author{Oleg Kim and Almut Beige}
\affiliation{The School of Physics and Astronomy, University of Leeds, Leeds LS2 9JT, United Kingdom}
\date{\today}

\begin{abstract}
Here we analyse cavity-mediated laser cooling for an experimental setup with an external trap which strongly confines the motion of a particle in the direction of the cavity axis. It is shown that the stationary state phonon number exhibits three sharp minima as a function of the atom-cavity detuning due to a direct atom-phonon-photon coupling term in the system Hamiltonian. These resonances have the same origin as the Mollow triplet in the resonance fluorescence of a laser-driven atomic system. It is shown that a laser-Rabi frequency-dependent atom-cavity detuning yields the lowest stationary state phonon number for a wide range of experimental parameters.
\end{abstract}

\maketitle

\section{Introduction}

Laser sideband cooling allows to cool single, strongly-confined atomic particles to very low temperatures \cite{WinelandDehmelt}. Its discovery opened the way for experiments which test the foundations of quantum physics and have applications ranging from quantum metrology to quantum computing \cite{RevMod}. Unfortunately, laser sideband cooling cannot be used to cool large numbers of trapped particles to very low temperatures. Moreover, laser sideband cooling cannot be used to cool particles with a very complex level structure, like molecules, very efficiently \cite{Lev}. Alternative cooling techniques therefore receive a lot of attention in the literature. First indications that cavity-mediated laser cooling allows to cool trapped particles to low temperatures were found in Paris already in 1995 \cite{Karine}. More systematic experimental studies of cavity-mediated laser cooling have subsequently been reported by several groups (cf.~Refs.~\cite{pinkse2,Rempe,vuletic,new,kimble,chap,Meschede,Wolke,Barrett}). 

\begin{figure}[t]
\center
\includegraphics[width=60mm]{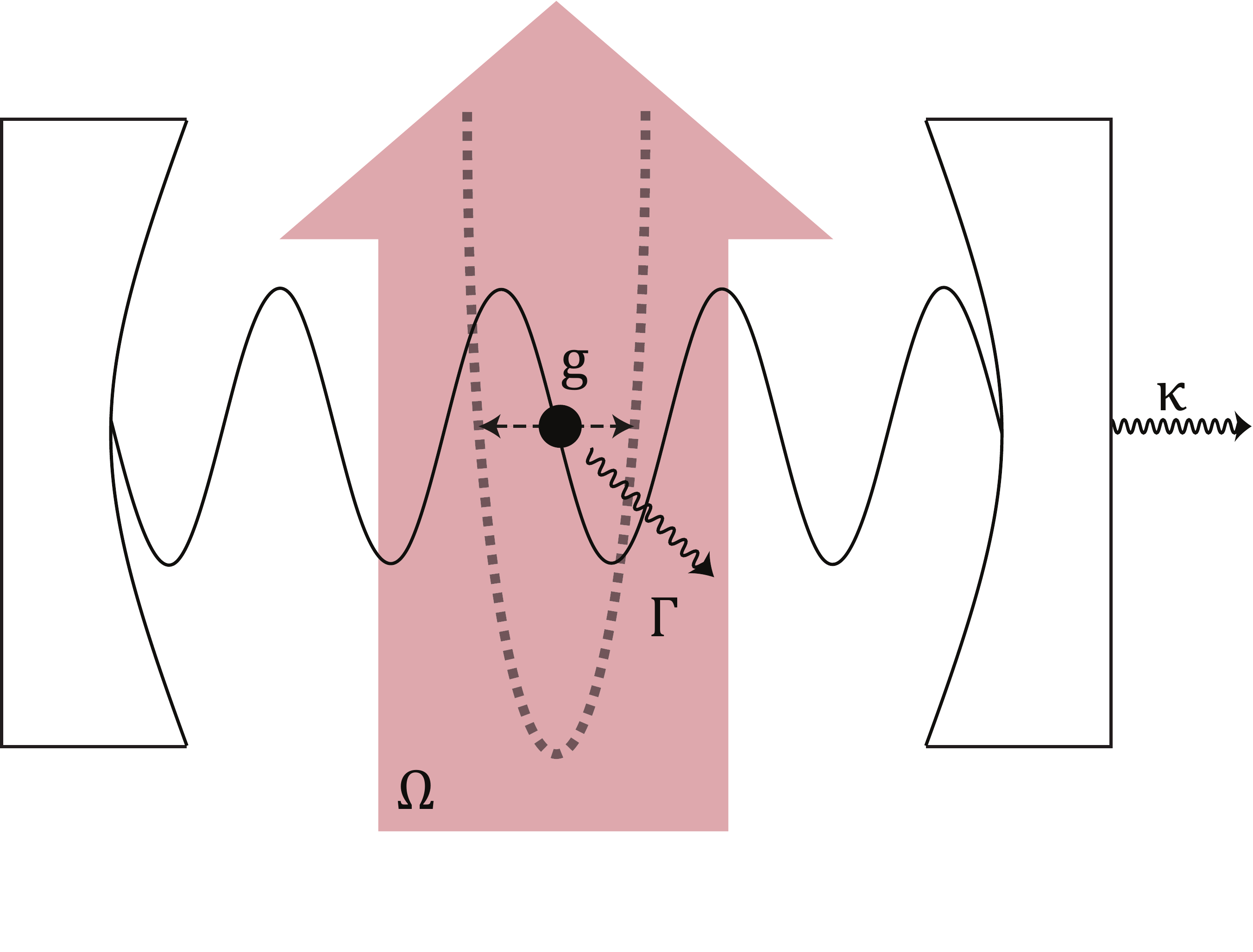} \\[-0.5cm]
\caption{(color online) Schematic view of the experimental setup. It consists of a resonantly driven atomic particle which is externally confined in the node of an optical cavity. Its motion is quantised in the direction of the cavity axis.} \label{setup}
\end{figure}

The theory of cavity-mediated laser cooling of free particles was first discussed in Refs.~\cite{lewen,lewen2}. Later, Ritsch and collaborators \cite{peter2,peter3} and others \cite{Murr,vuletic10,Robb} developed semiclassical theories to model cavity-mediated cooling processes. In 1993, Cirac {\em et al.}~\cite{Cirac2} introduced a master equation approach to analyse cavity-mediated laser cooling. Since the precision of calculations which are based on master equations is easier to control than the precision of semiclassical calculations, this approach has been used by many authors to show a close analogy between laser-sideband and cavity-mediated laser cooling \cite{Cirac4,cool,morigi,Morigi,Tony,Tony3}. In this paper we use the same methodology as in Refs.~\cite{Tony,Norah,Tony3} and analyse the cooling dynamics of the experimental setup in Fig.~\ref{setup} with the help of linear differential equations, so-called rate equations, for expectation values.

In the literature it is often assumed that the atom is trapped in a direction orthogonal to the cavity axis \cite{Cirac4,cool,Tony3}. Here we assume instead that an external trap confines the motion of a single particle in the direction of the cavity axis \footnote{This case is a special case of the general scenario considered in Refs.~\cite{morigi,Morigi}.}. As illustrated in Fig.~\ref{setup}, the particle should be placed in a node of the quantised standing wave cavity field mode. One way to achieve this is to drive the cavity and to create a strong optical lattice trapping potential. Alternatively, the particle could be held in a three-dimensional magneto optical trap. To initiate the cooling process, a laser field with Rabi frequency $\Omega$ drives the particle from the side. 

In cavity-mediated laser cooling, the reduction in the mean phonon number of the trapped particle is due to the continuous conversion of phonons into cavity photons. These phonons are permanently lost from the system when cavity photons leak out of the resonator. The result is a reduction of the kinetic energy of the trapped particle, ie.~cooling. One of the roles of the atomic particle in the cooling process is to facilitate the phonon-photon conversion which needs to be accompanied by certain electronic transitions. The purpose of the applied laser field is to populate certain atomic states which are involved in this process. Calculations which go beyond the scope of this paper have already shown that resonant laser driving of the atomic 0--1 transition supports the cooling process best. We therefore consider in the following the case of zero laser detuning.

As we shall see below, the stationary state phonon number $m_{\rm ss}$ of the experimental setup in Fig.~\ref{setup} exhibits three sharp minima as a function of the atom-cavity detuning $\delta$. The corresponding atom-cavity resonances have the same origin as the Mollow triplet in the resonance fluorescence of a laser-driven atomic system \cite{Mollow,Valle}. For relatively small spontaneous decay rates, they are simply given by 
\beq \label{sideband}
\delta_0 \equiv \nu ~& {\rm and} &~ \delta_\pm \equiv  \nu \pm \Omega \, .
\eeq
Although the experimental setup which we consider here has already been discussed in the literature \cite{morigi,Morigi}, the cooling potential of all three resonances has not yet been analysed in the literature. For a relatively wide range of experimental parameters, the previously unconsidered resonance $\delta = \delta_+$ yields a lower stationary state phonon number than the optimal detuning $\delta = \nu$ of laser sideband cooling \cite{RevMod}. Our calculations show that this cooling resonance is especially then of interest, when the spontaneous cavity decay rate $\kappa$ is relatively large or when the phonon frequency $\nu$ is relatively small.

There are five sections in this paper. In Section \ref{sec2}, we introduce the master equations for the description of the experimental setup shown in Fig.~\ref{setup}. We then use this equation to obtain a closed set of rate equations, ie.~linear differential equations for the time evolution of expectation values. These can be used to model the time evolution of the mean phonon number $m$ up to second order in the Lamb-Dicke parameter $\eta$. Before doing so, Section \ref{sec3} uses a simple argument to identify the relevant cooling and heating resonances. A detailed analysis of the cooling process with analytical and numerical results can be found in Section \ref{sec4}. Finally, we summarise our findings in Section \ref{conclusions}. 

\section{Theoretical model} \label{sec2}

In this section, we introduce a closed set of rate equations which can be used to analyse the dynamics of the experimental setup in Fig.~\ref{setup}. These are linear differential equations for the time evolution of expectation values, including one for the time evolution of the mean phonon number 
\begin{eqnarray}
m &\equiv & \langle b^\dagger b \rangle \, .
\end{eqnarray}
Additional expectation values are taken into account in order to obtain a closed set of rate equations. These accurately describe the time evolution of the system on a time scale proportional to $\eta^2$. 

\subsection{The time evolution of expectation values}

Let us now have a closer look at the Hamiltonian of the experimental setup shown in Fig.~\ref{setup}. In the usual dipole and rotating wave approximation, it equals 
\begin{eqnarray} \label{1}
H &=& \hbar \omega_0 \, \sigma^+ \sigma^- + \hbar \nu \, b^\dagger b + \hbar \omega_{\rm cav} \, c^\dagger c \nonumber \\
&& + \hbar g \, \sin \big( {\bf k}_{\rm cav} \cdot \textbf{r} \big) \, c \sigma^+ + {\rm H.c.} \nonumber \\
&& + {1 \over 2} \hbar \Omega \, \sigma^+ \, {\rm e}^{-{\rm i}\omega_0 t} + {\rm H.c.} 
\end{eqnarray}
Here $\hbar \omega_0$, $\hbar \nu$, and $\hbar \omega_{\rm cav}$ denote the energy difference between the atomic ground state $|0 \rangle$ and the excited state $|1 \rangle$, the free energy of a single phonon, and the free energy of a cavity photon. Moreover, $\sigma^+ \equiv |1 \rangle \langle 0|$ and $\sigma^- \equiv |0 \rangle \langle 1|$, while $b$ and $c$ are phonon and photon annihilation operators with bosonic commutator relations,
\begin{eqnarray}
[b, b^\dagger ] = [c, c^\dagger ] = 1 \, . 
\end{eqnarray}
The second line in Eq.~(\ref{1}) takes the atom-cavity interaction at the position ${\bf r}$ of the trapped particle into account. Here $g$ is the atom-cavity coupling constant and ${\bf k}_{\rm cav}$ is the wave vector of the cavity field. The third line describes the resonant driving of the particle with a laser with Rabi frequency $\Omega$ and frequency $\omega_0$.

In this paper, we assume an external trap which confines the motion of the particle in the direction of the cavity axis. We denote the trap centre by ${\bf R}$ and the displacement of the atom from the trap center by ${\bf x}$ such that its position ${\bf r}$ is given by ${\bf r} = {\bf R} + {\bf x}$. Considering the center of mass motion of the trapped particle quantised with the phonon annihilation operator $b$ from above yields 
\begin{eqnarray}
{\bf k}_{\rm cav} \cdot \textbf{x} &=& \eta (b + b^\dagger) \, . 
\end{eqnarray}
The Lamb-Dicke parameter $\eta $ in this equation is a measure for the strength of the trapping potential. Usually, one has 
\beq
\eta &\ll& 1\, . 
\eeq
For a wide range of particle positions ${\bf r}$, the atom-phonon-photon interaction is therefore only relatively weak. In order to maximise it, we assume in the following that ${\bf R}$ points at a node of the cavity field which implies ${\rm e}^{-{\rm i} {\bf k}_{\rm cav} \cdot \textbf{R}} = \pm 1$. Hence
\begin{eqnarray}\label{2.22d}
\sin \big( {\bf k}_{\rm cav} \cdot \textbf{r} \big) = \pm \eta \, (b + b^\dagger) + {\cal O}(\eta^3) \, . 
\end{eqnarray}
Substituting this into Eq.~(\ref{1}) and going into the interaction picture with respect to 
\beq
H_0 &=& \hbar \omega_0 ( \sigma^+ \sigma^- + c^\dagger c) \, , 
\eeq
we obtain the time-independent interaction Hamiltonian
\begin{eqnarray} \label{HI2first}
H_{\rm I} &=& \hbar \nu \, b^\dagger b + \hbar \delta \, c^\dagger c + {1 \over 2} \hbar \Omega \, (\sigma^- + \sigma^+) \nonumber \\
&& + \hbar \eta g \, (b + b^\dagger) (\sigma^+ c + \sigma^- c^\dagger ) + {\cal O}(\eta^3) 
\end{eqnarray}
in the usual Lamb-Dicke approximation. Here we ignored the minus sign in Eq.~(\ref{2.22d}), since this phase factor has no real physical consequences \cite{sign}. 

The main difference between laser-sideband \cite{WinelandDehmelt,RevMod} and cavity-mediated laser cooling is that, in the latter case, the atomic raising operator $\sigma^+$ in the cooling Hamiltonian is replaced by the cavity photon creation operator $c^\dagger$. Hence, the cooling efficiency depends strongly on the spontaneous cavity decay rate $\kappa$ and not only on the spontaneous atom decay rate $\Gamma$. To model this, spontaneous photon emission is in the following taken into account by the quantum optical master equation
\begin{eqnarray} \label{master}
\dot{\rho}_{\rm I} &=& - {{\rm i} \over \hbar} \left[H_{\rm I},\rho_{\rm I} \right] + {1 \over 2} \kappa \left(2 c \rho_{\rm I} c^\dagger - c^\dagger c \rho_{\rm I} - \rho_{\rm I} c^\dagger c \right) \nonumber \\
&& + {1 \over 2} \Gamma \left( 2 \sigma^- \rho_{\rm I} \sigma^+ - \sigma^+ \sigma^- \rho_{\rm I} - \rho_{\rm I} \sigma^+ \sigma^- \right) \, .
\end{eqnarray}
Using this equation, one can show that the time evolution of the expectation value of an arbitrary operator $A_{\rm I}$ in the interaction picture is given by 
\begin{eqnarray} \label{10}
\langle\dot{A_{\rm I}} \rangle &=& - {{\rm i} \over \hbar} \langle \left[A_{\rm I},H_{\rm I} \right] \rangle + {1 \over 2} \kappa \left\langle 2 c^\dagger A_{\rm I} c   - A_{\rm I} c^\dagger c  - c^\dagger c A_{\rm I} \right\rangle \nonumber \\
&& + {1 \over 2} \Gamma \left \langle 2 \sigma^+ A_{\rm I} \sigma^- - A_{\rm I} \sigma^+ \sigma^-  - \sigma^+ \sigma^- A_{\rm I} \right\rangle \, . ~~~
\end{eqnarray}
In the following, we use this equation to analyse the cooling process on a time scale proportional to $\eta^2$. 

\subsection{The relevant expectation values}

As we shall see below, in order to obtain a closed set of cooling equations, including one for the time evolution of the mean phonon number $m$, we need to consider the expectation values of certain mixed operators $X_{ijk}$ of the form 
\beq
X_{ijk} &\equiv & B_i \Sigma_j C_k 
\eeq
with the $B$, $\Sigma$, and $C$ operators defined such that
\beq
(B_0,\Sigma_0,C_0) &\equiv & (1,1,1) \, , \nonumber \\
(B_1,\Sigma_1,C_1) &\equiv & (b^\dagger b, \sigma^+\sigma^- , c^\dagger c) \, , \nonumber \\
(B_2, \Sigma_2, C_2) &\equiv & (b+b^\dagger, \sigma^- + \sigma^+ , c+c^\dagger ) \, , \nonumber \\
(B_3, \Sigma_3, C_3) &\equiv & {\rm i} (b-b^\dagger, \sigma^- - \sigma^+ , c-c^\dagger ) \, , \nonumber \\
(B_4,C_4) &\equiv & (b^2+b^{\dagger 2}, c^2+c^{\dagger 2}) \, , \nonumber \\
(B_5,C_5) &\equiv & {\rm i} (b^2 - b^{\dagger 2}, c^2 - c^{\dagger 2}) \, .
\end{eqnarray}
Using these operators, the Hamiltonian $H_{\rm I}$ in Eq.~(\ref{HI2first}) becomes
\begin{eqnarray} \label{HI4}
H_{\rm I} &=& \hbar \nu \, B_1 + \hbar \delta \, C_1 + {1 \over 2} \hbar \Omega \, \Sigma_2 \nonumber \\
&& + {1 \over 2} \hbar \eta g \, B_2 ( \Sigma_2 C_2 + \Sigma_3 C_3 ) \, .
\end{eqnarray}
In the following, we use this representation of the Hamiltonian, since the $X$ operators obey relatively simple commutator relations. Moreover, we denote their expectation values by 
\begin{eqnarray}
x_{ijk} & \equiv & \langle X_{ijk} \rangle \, . 
\end{eqnarray}
Since all the operators $X_{ijk}$ are Hermitian, the variables $x_{ijk}$ are all real.

\subsection{Time evolution of $m$ in first order in $\eta$} \label{etafirst}

Let us now have a closer look at the differential equations which describe the time evolution of $m$ and the $x_{ijk}$'s. Using Eqs.~(\ref{10}) and (\ref{HI4}) we find for example that the mean phonon number $m$ evolves according to
\beq \label{dotm}
\dot{m} &=& {1 \over 2} \eta g \, ( x_{322} + x_{333} ) \, .
\eeq
This means, if we are interested in the time evolution of $m$ on a time scale proportional to $\eta^n$, then we need to be able to calculate $x_{322}$ and $x_{333}$ up to order $n-1$ in $\eta$. In order to distinguish terms in different order in $\eta $ more easily, we adopt the notation
\beq
x &\equiv & x^{(0)} + x^{(1)} + ... 
\eeq
throughout the rest of this manuscript. The superscripts indicate the scaling of the respective contribution in $\eta$ to a certain variable $x$. 

First we have a look at the case $\eta = 0$. This means, we assume that there is no coupling between phonons, photons, and electrons. Hence the cavity remains in its vacuum state and 
\beq \label{6}
\langle C_k \rangle^{(0)} &\equiv & 0 
\eeq
for $k=1,...,5$. Analogously, by using again Eqs.~(\ref{10}) and (\ref{HI4}), one can show that 
\beq \label{y0}
\dot m^{(0)} &=& 0 \, . 
\eeq
This tells us that there is no cooling in zeroth order in $\eta$. Moreover we find that 
\beq \label{y0}
&& \hspace*{-0.4cm} \langle \dot B_2 \rangle^{(0)} = - \nu \, \langle B_3 \rangle^{(0)} \, , ~~
\langle \dot B_3 \rangle^{(0)} =\nu \, \langle B_2 \rangle^{(0)} \, , \nonumber \\ 
&& \hspace*{-0.4cm} \langle \dot B_4 \rangle^{(0)} = - 2 \nu \, \langle B_5 \rangle^{(0)} \, , ~~
\langle \dot B_5 \rangle^{(0)} = 2 \nu \, \langle B_4 \rangle^{(0)} \, . ~~~
\eeq
When solving these rate equations, we find that the phonon coherences $ \langle B_2 \rangle^{(0)}$ to $ \langle B_5 \rangle^{(0)}$ oscillate around zero on time scales given by the phonon frequency $\nu$. When analysing the cavity-mediated cooling process on a much longer time scale, the above phonon coherences can be adiabatically eliminated from the system dynamics. Setting their time derivatives in Eq.~(\ref{y0}) equal to zero yields
\beq \label{8}
\langle B_i \rangle^{(0)} &\equiv & 0 
\eeq
with $i=2,...,5$.  Later in Section \ref{sec4}, we use this equation to analyse the cooling dynamics of our system in more detail. As we shall see below, this approximation is well justified, since the cooling rate $\gamma_{\rm c}$ of the experimental setup which we consider here scales as $\eta^2$. 

After introducing the short hand notation $z_j \equiv \langle \Sigma_j \rangle^{(0)}$ for the electronic states of the trapped particle, one can show that 
\beq
&& \dot z_1 = {1 \over 2} \Omega \, z_3 - \Gamma \, z_1 \, , ~~ 
\dot z_2 = - {1 \over 2} \Gamma \, z_2 \, , \nonumber \\
&& \dot z_3 = \Omega \left(1 - 2 z_1 \right) - {1 \over 2} \Gamma \, z_3 
\eeq
in zeroth order in $\eta$. These expectation values reach a stationary state relatively quickly. When analysing process on the time scale given by the cooling rate $\gamma_{\rm c}$, these too can be approximated by their stationary state solutions. Doing so, we assume in the following that
\beq \label{zs}
(z_1,z_2,z_3) &=& \left( {\Omega^2 \over \Gamma^2 + 2 \Omega^2} , 0, {2 \Gamma \Omega \over \Gamma^2 + 2 \Omega^2} \right) \, .
\eeq
Before using these results to derive an effective cooling equation for the mean phonon number $m$, we notice that
\beq \label{11}
x_{ijk}^{(0)} &=& \langle B_i \rangle^{(0)} \langle \Sigma_j \rangle^{(0)} \langle C_k \rangle^{(0)}
\eeq
when $\eta = 0$, since all three subsystems evolve independently in this case. Hence $x_{322}^{(0)} = x_{333}^{(0)} = 0$. Substituting this into Eq.~(\ref{dotm}) yields 
\beq
\dot m^{(1)} &=& 0 \, . 
\eeq
To obtain a non-zero time derivative of $m$, we need to calculate $x_{322}$ and $x_{333}$, at least, up to first order in $\eta$.

\subsection{Time evolution of $m$ in second order in $\eta$} \label{second}

Taking the results in Eqs.~(\ref{6}), (\ref{8}), (\ref{zs}), and (\ref{11}) into account, it is now possible to obtain a closed set of cooling equations for the coherences $x_{322}^{(1)}$ and $x_{333}^{(1)}$. These are given by the linear differential equations
\beq \label{42}
\dot{x}_{202}^{(1)} &=& - \nu \, x_{302}^{(1)} -\delta \, x_{203}^{(1)} - \eta g \, \left(1 + 2 m^{(0)} \right) z_3 - {1 \over 2} \gamma_0 \, x_{202}^{(1)} \, , \nonumber \\
\dot{x}_{203}^{(1)} &=& -\nu \, x_{303}^{(1)} +\delta \, x_{202}^{(1)} + \eta g \, \left(1 + 2 m^{(0)} \right) z_2 - {1 \over 2} \gamma_0 \, x_{203}^{(1)} \nonumber  \\
\dot{x}_{302}^{(1)} &=& \nu \, x_{202}^{(1)} -\delta \, x_{303}^{(1)} + \eta g \, z_2 - {1 \over 2} \gamma_0 \,  x_{302}^{(1)}  \nonumber  \\
\dot{x}_{303}^{(1)} &=& \nu \, x_{203}^{(1)} +\delta \, x_{302}^{(1)} + \eta g \, z_3 - {1 \over 2} \gamma_0 \, x_{303}^{(1)} \, ,
\eeq 
and
\beq \label{ones}
\dot{x}_{212}^{(1)} &=& -\nu \, x_{312}^{(1)} -\delta \, x_{213}^{(1)} + {1 \over 2} \Omega \, x_{232}^{(1)} - {1 \over 2} \gamma_2 \, x_{212}^{(1)} \, , ~~~ \nonumber \\
\dot{x}_{213}^{(1)} &=&-\nu \, x_{313}^{(1)} +\delta \, x_{212}^{(1)} +{1 \over 2} \Omega \, x_{233}^{(1)} - {1 \over 2} \gamma_2 \, x_{213}^{(1)} \, , \nonumber \\
\dot{x}_{312}&=&\nu \, x_{212}^{(1)} -\delta \, x_{313}^{(1)} +{1 \over 2} \Omega \, x_{332}^{(1)} - {1 \over 2} \gamma_2 \, x_{312}^{(1)} \, , \nonumber \\
\dot{x}_{313}^{(1)} &=&\nu \, x_{213}^{(1)} +\delta \, x_{312}^{(1)} + {1 \over 2} \Omega \, x_{333}^{(1)} - {1 \over 2} \gamma_2 \, x_{313}^{(1)} \, ,
\eeq
where $m^{(0)}$ denotes the zeroth order contribution of the mean phonon number $m$. Moreover, one can show that
\beq \label{twos}
\dot{x}_{222}^{(1)} &=&-\nu \,  x_{322}^{(1)} -\delta \,  x_{223}^{(1)}  - {1 \over 2} \gamma_1 \,  x_{222}^{(1)} \, , \nonumber \\
\dot{x}_{223}^{(1)} &=&-\nu \,  x_{323}^{(1)} +\delta \,  x_{222}^{(1)} + 2 \eta g \, \left(1 + 2 m^{(0)} \right) z_1 - {1 \over 2} \gamma_1 \,  x_{223}^{(1)} \, , \nonumber \\
\dot{x}_{322}^{(1)} &=&\nu \,  x_{222}^{(1)} -\delta \,  x_{323}^{(1)}  + 2 \eta g \, z_1 - {1 \over 2} \gamma_1 \,  x_{322}^{(1)} \, , \nonumber \\
\dot{x}_{323}^{(1)} &=&\nu \,  x_{223}^{(1)} +\delta \,  x_{322}^{(1)}  - {1 \over 2} \gamma_1 \,  x_{323}^{(1)} \, , 
\eeq
and
\beq \label{threes}
\dot{x}_{232}^{(1)} &=&-\nu \,  x_{332}^{(1)} -\delta \,  x_{233}^{(1)} +\Omega \left(x_{202}^{(1)} -2x_{212}^{(1)} \right) \nonumber \\ 
&& -2 \eta g \, \left(1 + 2m^{(0)} \right) z_1 - {1 \over 2} \gamma_1 \,  x_{232}^{(1)} \, , \nonumber \\
\dot{x}_{233}^{(1)} &=&-\nu \,  x_{333}^{(1)} +\delta \,  x_{232}^{(1)} +\Omega \left(x_{203}^{(1)} -2x_{213}^{(1)} \right) - {1 \over 2} \gamma_1 \,  x_{233}^{(1)} \, , \nonumber \\
\dot{x}_{332}^{(1)} &=&\nu \,  x_{232}^{(1)} -\delta \,  x_{333}^{(1)} + \Omega \left(x_{302}^{(1)} -2x_{312}^{(1)} \right) - {1 \over 2} \gamma_1 \,  x_{332}^{(1)} \, , \nonumber \\
\dot{x}_{333}^{(1)} &=&\nu \,  x_{233}^{(1)} +\delta \,  x_{332}^{(1)} + \Omega \left(x_{303}^{(1)} -2x_{313}^{(1)} \right) +2 \eta g \, z_1 \nonumber \\ 
&& - {1 \over 2} \gamma_1 \,  x_{333}^{(1)} \, .
\eeq
Here the effective spontaneous decay rates $\gamma_{\rm n}$ are defined such that
\beq \label{gamman}
\gamma_n &\equiv & \kappa + n \, \Gamma \, .
\eeq 
As we shall see below (cf.~Eq.~(\ref{mdot})), these equations indeed constitute a closed set of rate equations when combined with the differential equation for the time evolution of the mean phonon number $m$ in second order in $\eta$.

\section{Expected cooling and heating resonances} \label{sec3}

Phonons have no spontaneous decay rate. To initiate the cooling process, it is therefore important to convert them into particles with a non-zero spontaneous decay rate, like cavity photons. One of the roles of the atomic particle is to facilitate this conversion. By changing its electronic state, the atomic particle supports the conversion of a phonon into a cavity photon. When the photon subsequently leaks out of the cavity, a phonon is permanently lost which implies cooling. In order to make the cooling process as efficient as possible, the detunings of the experimental setup in Fig.~\ref{setup} should be adjusted such that cooling transitions become resonant. Moreover, all heating transitions should be as off-resonant as possible. For the experimental setup which we consider here, this means that at least some of the $bc^\dagger$-terms in the Hamiltonian need to be in resonance, while resonance of $b^\dagger c^\dagger$ terms should be avoided.

In order to identify the relevant cooling and heating resonances and to get more insight into the dynamics induced by the Hamiltonian $H_{\rm I}$, we now change into a dressed state picture. To do so, we diagonalise the laser driving term, ie.~the atomic operator $\sigma_{\rm x} = \sigma^- + \sigma^+$, in Eq.~(\ref{HI2}). The eigenvalues and eigenvectors of $\sigma_{\rm x}$ are $\lambda_\pm = \pm 1$ and 
\beq
|\lambda_\pm \rangle &=& {1 \over \sqrt{2}} \big( |0 \rangle \pm |1 \rangle \big) \, , 
\eeq
respectively. Using this notation, we find that
\beq
\sigma^\pm = {1 \over 2} \big( \,  |\lambda_+ \rangle \langle \lambda_+ | - |\lambda_- \rangle \langle \lambda_- | \pm |\lambda_+ \rangle \langle \lambda_- | \mp |\lambda_- \rangle \langle \lambda_+ | \, \big) \, . \nonumber \\
\eeq
Consequently, the Hamiltonian $H_{\rm I}$ in Eq.~(\ref{HI2first}) can be written as
\begin{eqnarray} \label{HI2}
H_{\rm I} &=& \hbar \nu \, b^\dagger b + \hbar \delta \, c^\dagger c + {1 \over 2} \hbar \Omega \left( |\lambda_+ \rangle \langle \lambda_+ | -  |\lambda_- \rangle \langle \lambda_- | \right) \nonumber \\
&& + {1 \over 2} \hbar \eta g \, (b + b^\dagger) (c + c^\dagger ) \big( \, |\lambda_+ \rangle \langle \lambda_+ | - |\lambda_- \rangle \langle \lambda_- | \, \big) \, , \nonumber \\
&& + {1 \over 2} \hbar \eta g \, (b + b^\dagger) (c - c^\dagger ) \big( \, |\lambda_+ \rangle \langle \lambda_- | - {\rm H.c.} \, \big) \, .
\end{eqnarray}
To remove all the terms in the first line of this equation from $H_{\rm I}$, we now go into a further interaction picture and obtain the interaction Hamiltonian $\tilde H_{\rm I}$,
\begin{eqnarray} \label{HI3}
\tilde H_{\rm I} &=& {1 \over 2} \hbar \eta g \, \left[ {\rm e}^{-{\rm i} (\delta + \nu) t} \, b c + {\rm e}^{- {\rm i} (\delta - \nu) t} \, b c^\dagger + {\rm H.c.} \right] \nonumber \\
&& \hspace*{1.5cm} \times \big( \, |\lambda_+ \rangle \langle \lambda_+ | - |\lambda_- \rangle \langle \lambda_- | \, \big) \nonumber \\
&& + {1 \over 2} \hbar \eta g \, \left[ {\rm e}^{-{\rm i} (\delta + \nu) t} \, b c - {\rm e}^{- {\rm i} (\delta - \nu) t} \, b c^\dagger - {\rm H.c.} \right] ~~ \nonumber \\
&& \hspace*{1.5cm} \times \big( \, {\rm e}^{{\rm i} \Omega t} \, |\lambda_+ \rangle \langle \lambda_- | - {\rm H.c.} \, \big) \, .
\end{eqnarray}
To assure that at least one of the $b c^\dagger$ terms in the above Hamiltonian becomes time-independent, the atom-cavity detuning $\delta$ needs to equal one of the three detunings $\delta_0$ and $\delta_\pm$ in Eq.~(\ref{sideband}). These three resonances are the three cooling resonances of the cooling process which we consider here. Moreover, all heating terms, ie.~all $b^\dagger c^\dagger$ terms, should oscillate rapidly in time. This means, the atom-cavity detuning $\delta$ should stay away as far as possible from the three detunings 
\beq \label{heating}
\mu_0 \equiv - \nu ~& {\rm and} &~ \mu_\pm \equiv - \nu \pm \Omega \, .
\eeq
These are the three heating resonances of the cooling process. One can easily check that the distance between any neighboring cooling or heating resonances equals the laser Rabi frequency $\Omega$, ie.~$|\delta_0 - \delta_\pm|$ and $|\mu_0 - \mu_\pm|$. The same applies for the resonances of a laser-driven atomic two-level system inside a quantised field \cite{Mollow}. This means, the $\delta$ resonances in Eq.~(\ref{sideband}) and the $\mu$ resonances in Eq.~(\ref{heating}), respectively, form a so-called Mollow triplet.

\section{Detailed analysis of the cooling process} \label{sec4}

The discussion in the previous section tells us for which atom-cavity detunings we can expect relatively efficient cooling of the trapped particle, as long as the spontaneous decay rates $\kappa$ and $\Gamma$ remain relatively small. To learn more about the cooling process and to study the effect of relatively large spontaneous decay rates on the cooling process, we now analyse the above described cooling process in more detail. To do so, we assume that the atom-phonon-photon interaction constant $\eta g$ is either much smaller than the atom-cavity detuning $\delta$, or much smaller than the cavity decay rate $\kappa$, or much smaller than the phonon frequency $\nu$,
\begin{eqnarray} \label{condi}
\eta g &\ll & \delta, \, \kappa , ~ {\rm or} ~ \nu \, .
\end{eqnarray}
The above discussion and a closer look at the above rate equations, especially Eq.~(\ref{dotm}), shows that the mean phonon number $m$ evolves on the very slow time scale given by $\eta g$. Looking at Eqs.~(\ref{42})--(\ref{threes}), we moreover find that the relevant $x$-coherences evolve on a time scale given by the largest one of the three frequencies $\delta$, $\kappa$, and $\nu$. In other words, Eq.~(\ref{condi}) guarantees that the mean phonon number $m$ evolves on a much slower time scale than the relevant coherences $x_{ijk}^{(1)}$. In other words, the time evolution of the mean phonon number is much slower than the inner dynamics of the atom-cavity-phonon system. The following detailed analysis confirms this assumption. No conditions, other than Eq.~(\ref{condi}), need to be imposed for the following calculations to apply.

\subsection{An effective cooling equation}

The above condition allows us to calculate the coherences $x_{ijk}^{(1)}$ via an adiabatic elimination. Doing so and setting for example the time derivatives of the coherences $x_{ijk}^{(1)}$ with $j=2$ in Eq.~(\ref{twos}) equal to zero, we obtain an expression for $x_{322}^{(1)}$. To calculate $x_{333}^{(1)} $, the remaining 12 rate equations in Eqs.~(\ref{42}), (\ref{ones}), and (\ref{threes}) have to be taken into account. Setting them equal to zero and substituting the resulting expressions for $x_{322}^{(1)}$ and $x_{333}^{(1)}$ into Eq.~(\ref{dotm}), we finally find that 
\begin{widetext}
\beq \label{mdot}
\dot m^{(2)} &=& {2 \eta^2 g^2 \Omega^2 \over \Gamma^2 + 2 \Omega^2} \left\{ {\gamma_1 \over \gamma_1^2 + \xi_+^2} + {\left( \gamma_0 \gamma_1 \gamma_2 +  \gamma_{-1} \xi_+^2 \right) \left( \gamma_2^2 + \xi_+^2 \right) + 4  \Omega^2 \left( \gamma_0 \gamma_2^2 + \gamma_4 \xi_+^2 \right) \over \left(\gamma_0^2 + \xi_+^2 \right) \left[ \left(\gamma_1^2 + \xi_+^2 \right) \left(\gamma_2^2 + \xi_+^2 \right) + 8 \Omega^2 \left(\gamma_1 \gamma_2 - \xi_+^2 \right) + 16 \Omega^4 \right]}  \right\} \left(1+ m^{(0)} \right) \nonumber \\
&& - {2 \eta^2 g^2 \Omega^2 \over \Gamma^2 + 2 \Omega^2} \left\{ {\gamma_1 \over \gamma_1^2 + \xi_-^2} 
+ {\left( \gamma_0 \gamma_1 \gamma_2 + \gamma_{-1} \xi_-^2 \right) \left( \gamma_2^2 + \xi_-^2 \right) + 4 \Omega^2 \left( \gamma_0 \gamma_2^2 + \gamma_4 \xi_-^2 \right) \over \left(\gamma_0^2 + \xi_-^2 \right) \left[ \left(\gamma_1^2 + \xi_-^2 \right) \left( \gamma_2^2 + \xi_-^2 \right) + 8 \Omega^2 \left( \gamma_1 \gamma_2 - \xi_-^2 \right) + 16 \Omega^4 \right]} \right\} \, m^{(0)}  
\eeq
\end{widetext}
with the parameter $\xi_\pm$ defined as 
\begin{eqnarray}
\xi_\pm \equiv 2 (\delta \pm \nu) 
\end{eqnarray} 
and with the $\gamma_n$ defined as in Eq.~(\ref{gamman}). 

Setting the time derivative $\dot m^{(2)}$ in Eq.~(\ref{mdot}) equal to zero yields an analytical expression for the stationary state phonon number $m_{\rm ss}$ of the proposed cooling process in zeroth order in $\eta$. Unfortunately, this expression is relatively complex and looking at it does not yield much insight into the considered cavity-mediated laser cooling process. In the following, we therefore only notice that the above cooling equation is of the general form
\beq \label{moremdot}
\dot m^{(2)} &=& - \gamma_{\rm c}^{(2)} \, m^{(0)} + c^{(2)} \, ,
\eeq
where $\gamma_{\rm c}^{(2)}$ is an effective cooling rate and where $c^{(2)}$ is a constant. Both $\gamma_{\rm c}^{(2)}$ and $c^{(2)}$ scale as $\eta^2$. In the following, we discuss the dependence of $\gamma_{\rm c}^{(2)}$ and of the stationary state phonon number $m_{\rm ss}^{(0)}$,
\beq
m_{\rm ss}^{(0)} = c^{(2)}/\gamma_{\rm c}^{(2)} \, ,
\eeq
on the different experimental parameters of the atom-cavity system in Fig.~\ref{setup}.

\subsection{Confirmation of the relevant cooling and heating resonances}

Before doing so, let us have a closer look at Eq.~(\ref{mdot}). Suppose that the laser driving is so weak that all the $\Omega^2$ terms in Eq.~(\ref{mdot}) become negligible. In this case, we find that
\beq \label{theusual}
m_{\rm ss}^{(0)} &=& {\kappa^2 + 4 (\delta - \nu)^2 \over 16 \delta \nu} \, .
\eeq
This stationary state phonon number is exactly the same as $m_{\rm ss}$ for laser sideband cooling of a trapped particle in free space \cite{WinelandDehmelt,RevMod} but with $\Gamma$ replaced by $\kappa$. For relatively small cavity decay rates $\kappa$, it assumes its minimum when $\delta = \delta_0$ with $\delta_0$ defined as in Eq.~(\ref{sideband}). Looking only at the case of weak laser driving, one might indeed conclude that there is only a single cooling resonance and a very close analogy between laser sideband and cavity-mediated laser cooling. Instead, this paper illustrates that atom-cavity-phonon systems can exhibit a much richer inner dynamics than systems with only atom-phonon interactions. 

\begin{figure}[t]
\center
\includegraphics[width=90mm]{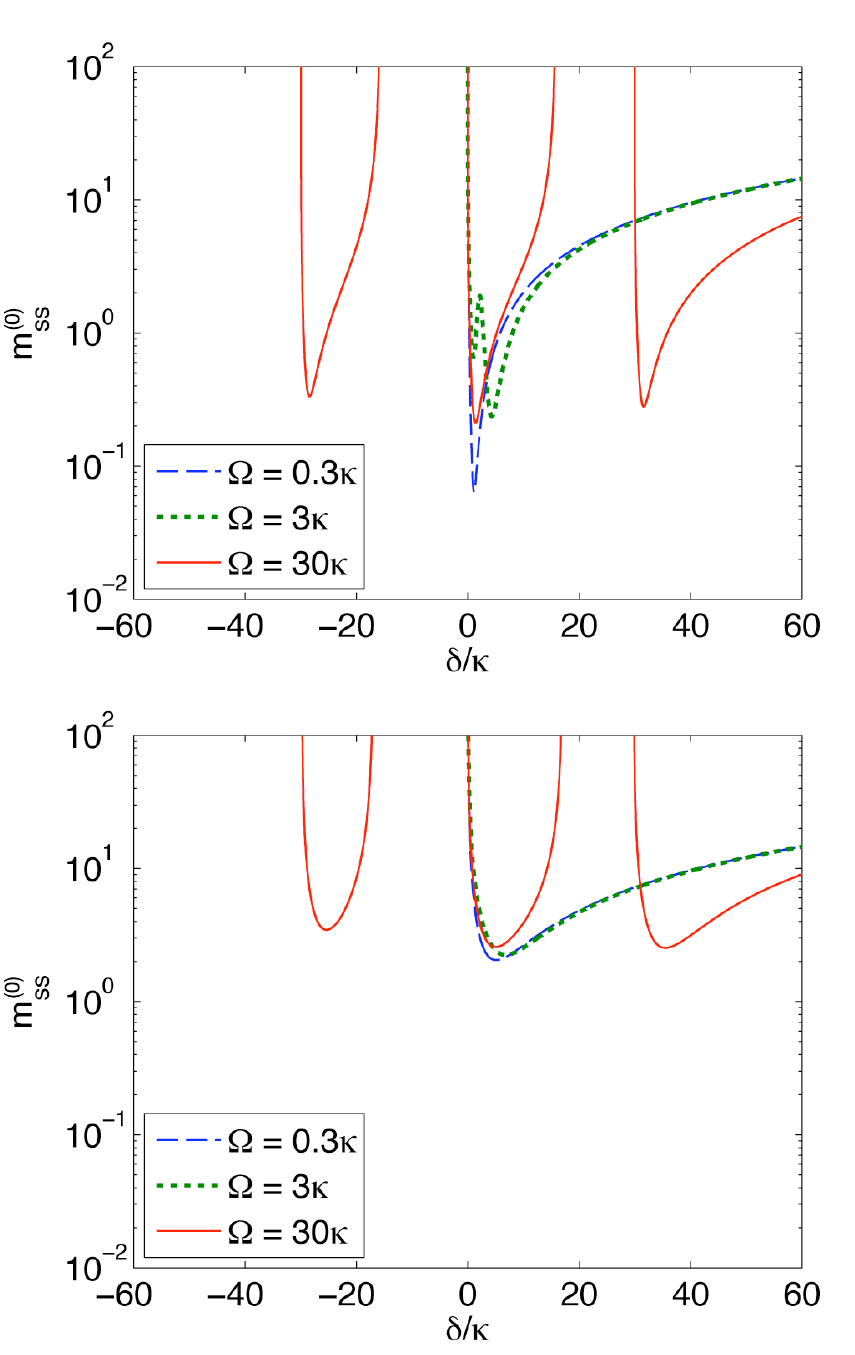}
\caption{(color online) Logarithmic plot of the stationary state phonon number $m_{\rm ss}^{(0)}$ as a function of the atom-cavity detuning $\delta$ for three different Rabi frequencies $\Omega$ and $\nu = \Gamma$, while $\kappa = \Gamma$ (upper figure) and $\kappa = 10 \, \Gamma$ (lower figure). This figure has been obtained from Eq.~(\ref{mdot}) by setting $\dot m^{(2)}$ equal to zero and clearly illustrates the presence of the cooling and heating resonances which we identified in Eqs.~(\ref{sideband}) and (\ref{heating}).} \label{fig2}
\end{figure}

\begin{figure}[t]
\center
\includegraphics[width=90mm]{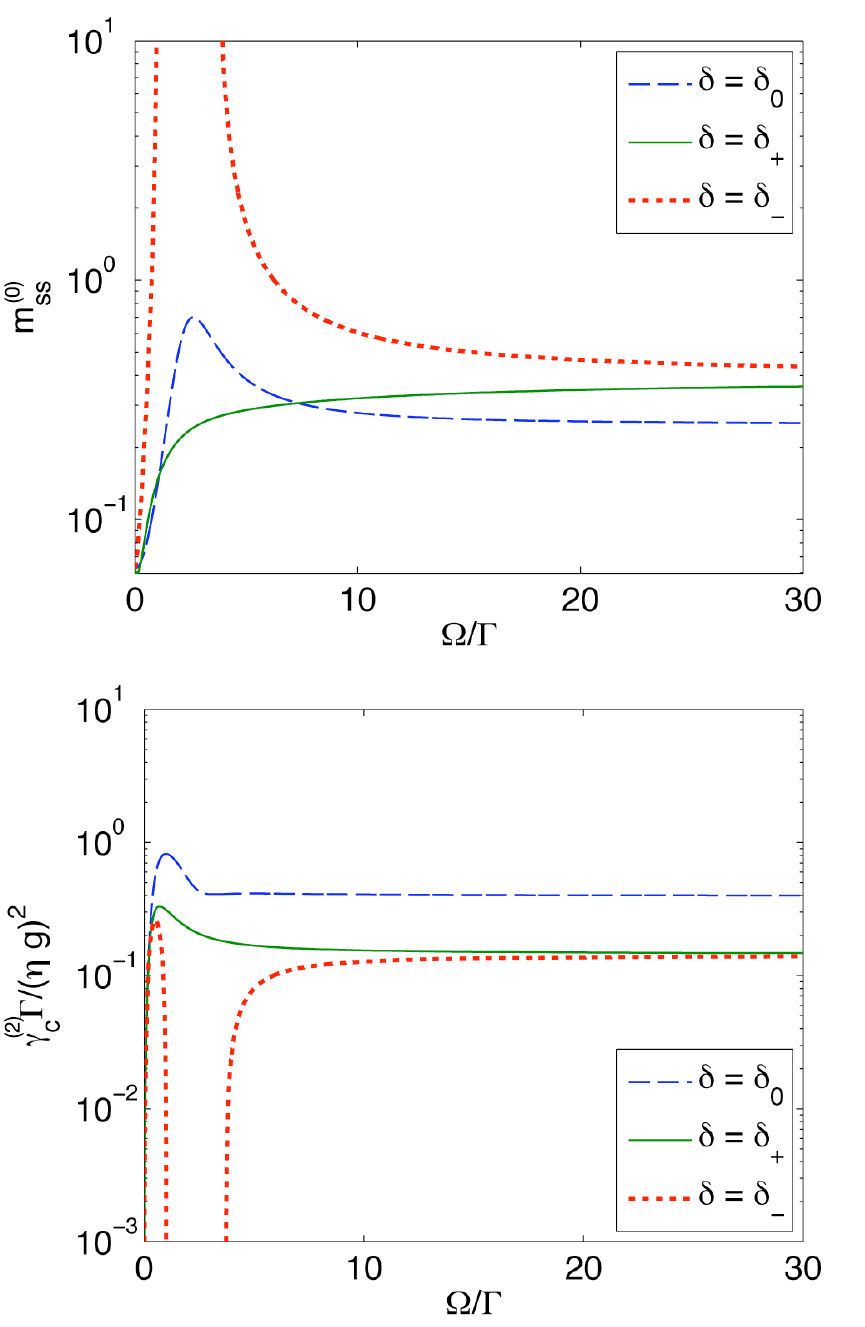}
\caption{(color online) Logarithmic plot of the stationary state phonon number $m_{\rm ss}^{(0)}$ and the cooling rate $\gamma_{\rm c}^{(2)}$ as a function of the laser Rabi frequency $\Omega$, while $\nu = \kappa = \Gamma$.} \label{fig4}
\end{figure}

Another interesting parameter regime is the one where $\Omega , \, \xi_{\pm} \gg \kappa \, , \Gamma$. In this case, Eq.~(\ref{mdot}) simplifies to
\beq
\dot m^{(2)} &=& \eta^2 g^2 \left\{ {\gamma_1 \over \xi_+^2} + {\gamma_{-1} \xi_+^2 + 4 \gamma_4 \Omega^2 \over \xi_+^4 - 8 \Omega^2 \xi_+^2 + 16 \Omega^4}  \right\} \left(1+ m^{(0)} \right) ~~~ \nonumber \\
&& - \eta^2 g^2 \left\{ {\gamma_1 \over \xi_-^2} + {\gamma_{-1} \xi_-^2 + 4 \gamma_4 \Omega^2 \over \xi_-^4 - 8 \Omega^2 \xi_-^2 + 16 \Omega^4} \right\} \, m^{(0)}  \, .
\eeq
The corresponding stationary state phonon number $m_{\rm ss}^{(0)}$ equals zero when $\xi_-^2 = 4 \Omega^2$, ie.~when $\delta$ equals either $\delta_-$ or $\delta_+$ in Eq.~(\ref{sideband}). This simple analysis confirms the presence of the two additional laser-Rabi frequency dependent cooling resonances $\delta_\pm$. However, notice that the above constraint $\xi_- \gg \kappa \, , \Gamma$ excludes the case where $\delta = \nu$. Hence this simple calculation returns only two of the three cooling resonances. 

We now return to Eq.~(\ref{mdot}) and use it to calculate the stationary state phonon number $m_{\rm ss}^{(0)}$ for the experimental setup in Fig.~\ref{setup} for concrete experimental parameters. Fig.~\ref{fig2} shows $m_{\rm ss}^{(0)}$ as a function of the atom-cavity detuning $\delta$ for a relatively wide range of parameters. To illustrate that the predictions in Section \ref{sec3} apply, even for relatively large spontaneous decay rates, we choose $\kappa$ and $\Gamma$ to be of about the same order of magnitude as the phonon frequency $\nu$ and the atom-cavity detuning $\delta$. For relatively large laser Rabi frequencies $\Omega$, we indeed observe three distinct cooling resonances with sharp local minima of the stationary state phonon number $m_{\rm ss}$. These are the atom-cavity detunings $\delta_0$ and $\delta_\pm$ which we defined in Eq.~(\ref{sideband}). In contrast to this and in good agreement with the discussion in Section \ref{sec3}, the stationary state phonon number $m_{\rm ss}^{(0)}$ increases significantly, when $\delta$ approaches one of the three heating resonances $\mu_0$ and $\mu_\pm$ in Eq.~(\ref{heating}). Only, when $\Omega$ becomes much smaller than $\nu$, then the cooling resonances and the heating resonances, respectively, become all the same. In this case, cooling occurs only for $\delta = \nu$ and extreme heating occurs for $\delta = - \nu$.

\begin{figure}[t]
\center
\includegraphics[width=90mm]{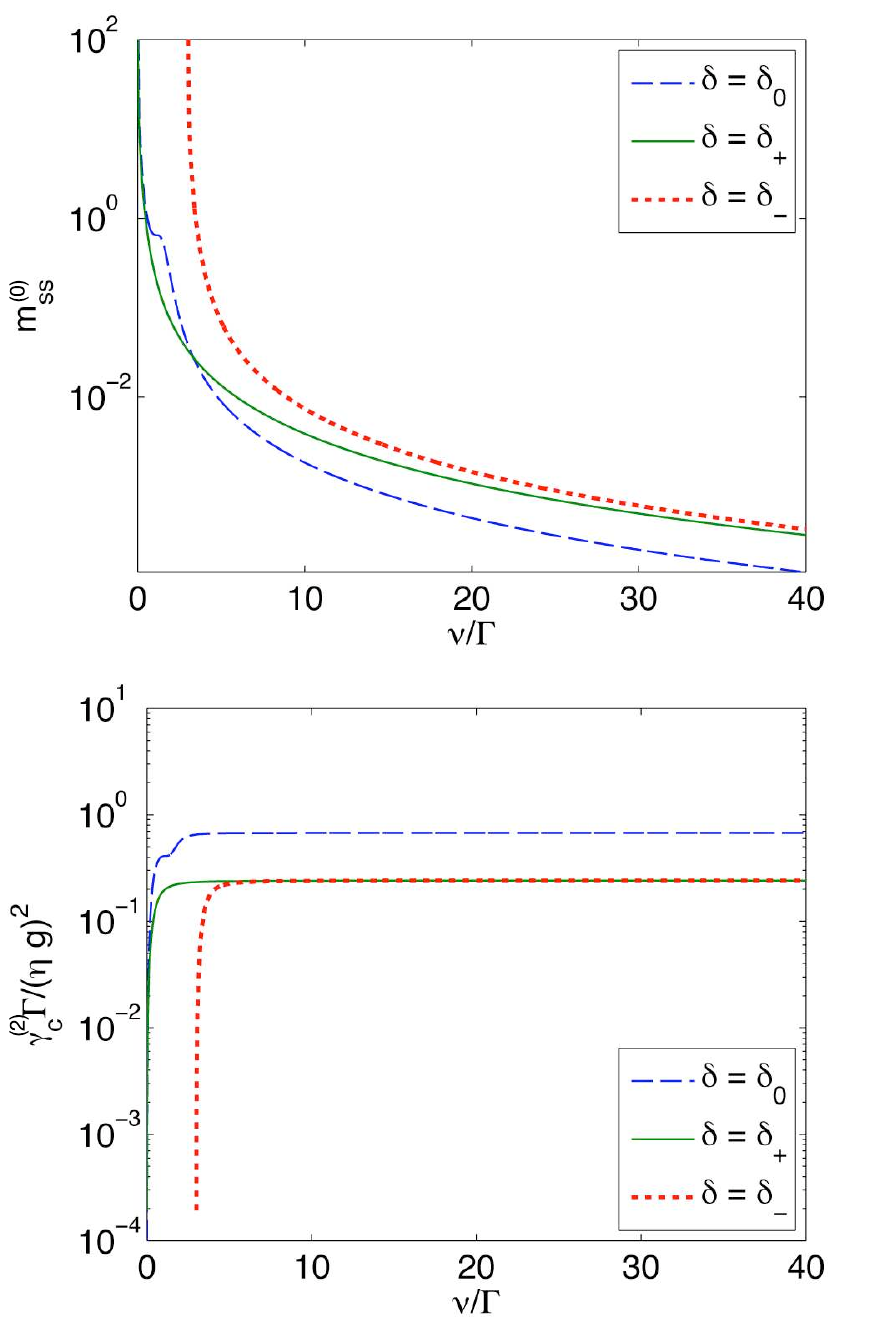}
\caption{(color online) Logarithmic plot of the stationary state phonon number $m_{\rm ss}^{(0)}$ and the cooling rate $\gamma_{\rm c}^{(2)}$ as a function of the phonon frequency $\nu$ for $\Omega=3 \, \Gamma$ and $\kappa = \Gamma$.} \label{fig5}
\end{figure}

\begin{figure}[t]
\center
\includegraphics[width=90mm]{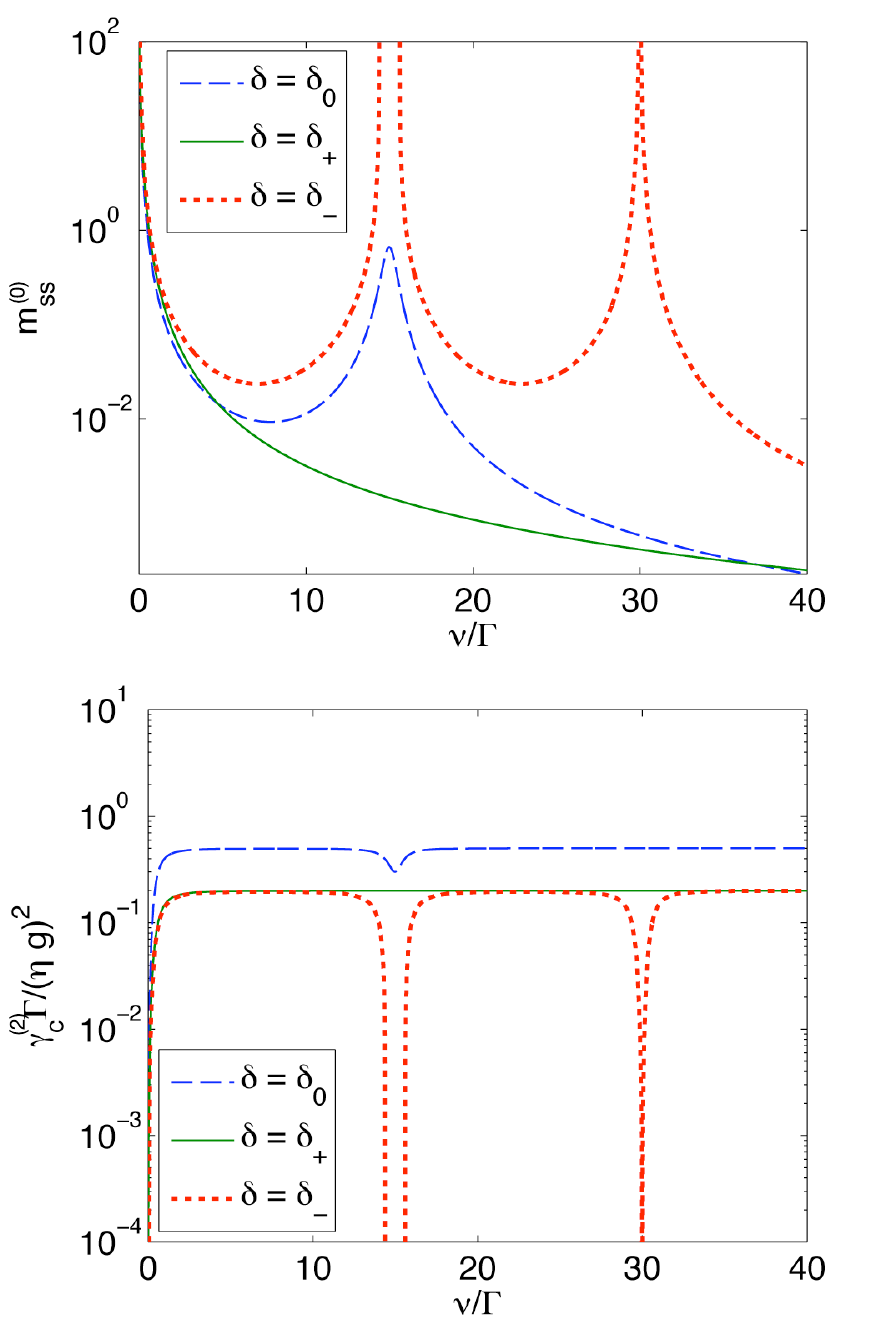}
\caption{(color online) Logarithmic plot of the stationary state phonon number $m_{\rm ss}^{(0)}$ and the cooling rate $\gamma_{\rm c}^{(2)}$ as a function of the phonon frequency $\nu$ for $\Omega=30 \, \Gamma$ and $\kappa = \Gamma$.} \label{fig6}
\end{figure}

\subsection{A comparison of the three cooling resonances}

To find out how to best cool a trapped particle when using the experimental setup in Fig.~\ref{setup}, we now compare the stationary state phonon numbers $m_{\rm ss}^{(0)}$ and the effective cooling rates $\gamma_{\rm c}^{(2)}$ of the three cooling resonances $\delta_0$ and $\delta_\pm$ with each other. When comparing the expressions for $\dot m^{(2)}$ in Eqs.~(\ref{mdot}) and (\ref{moremdot}), we find that $\gamma_{\rm c}^{(2)}$ becomes independent of $\eta$ and $g$ when dividing it by $(\eta g)^2$. The following results therefore apply for any values of these two parameters, as long as they fulfill the condition which we specified in Eq.~(\ref{condi}). 

\begin{figure}[t]
\center
\includegraphics[width=90mm]{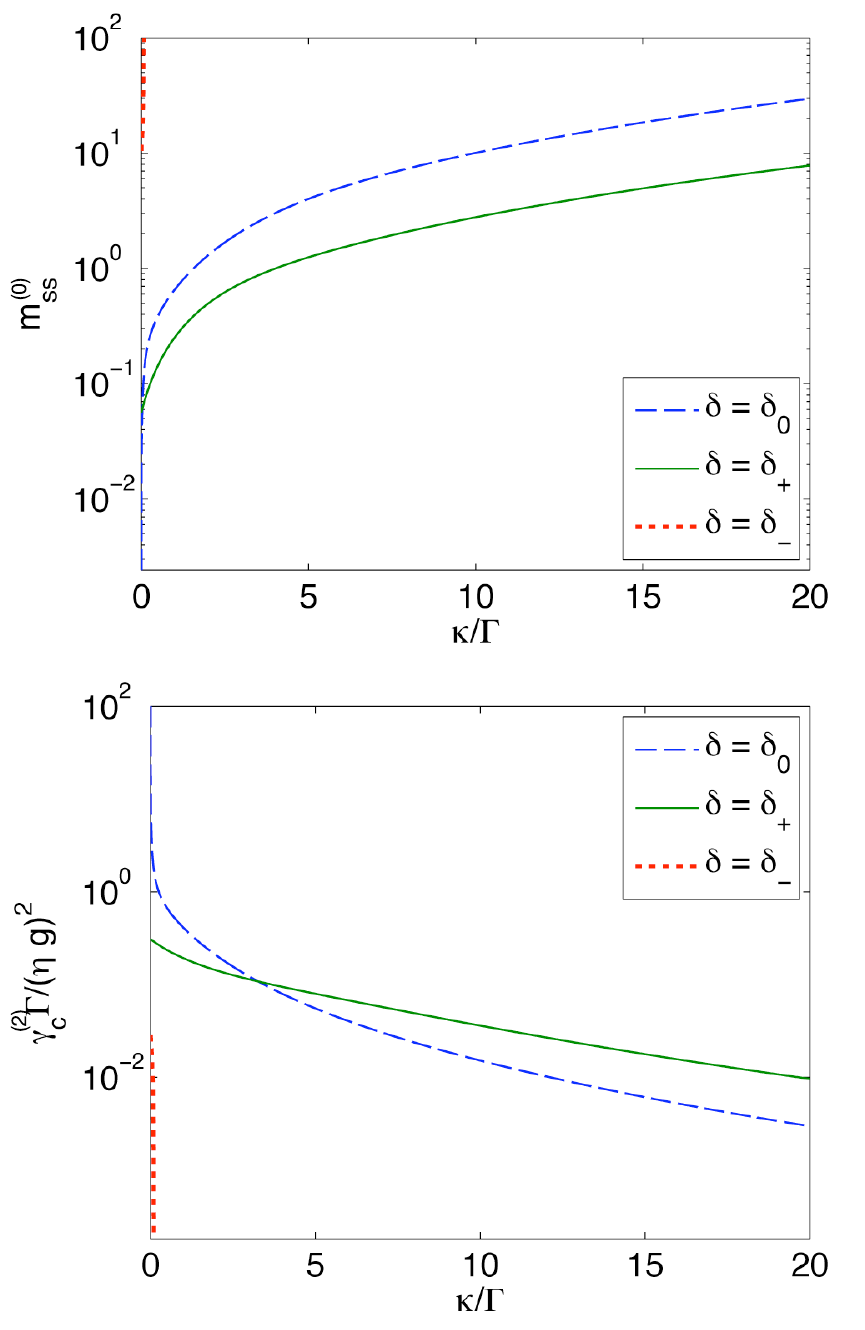}
\caption{(color online) Logarithmic plot of the stationary state phonon number $m_{\rm ss}^{(0)}$ and the cooling rate $\gamma_{\rm c}^{(2)}$ as a function of the spontaneous cavity decay rate $\kappa$ for $\Omega=3 \, \Gamma$ and $\nu = \Gamma$.} \label{fig7}
\end{figure}

\begin{figure}[t]
\center
\includegraphics[width=90mm]{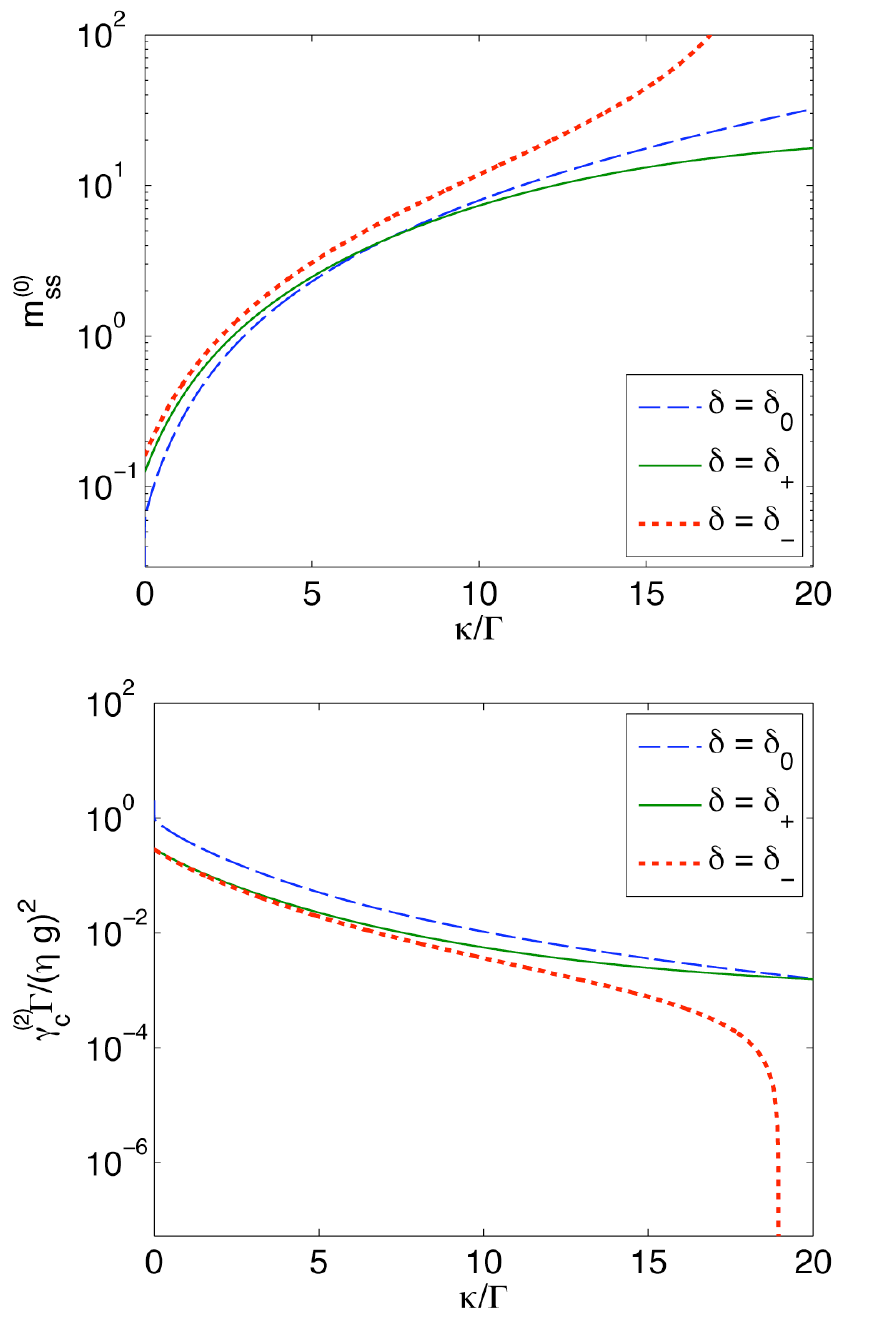}
\caption{(color online) Logarithmic plot of the stationary state phonon number $m_{\rm ss}^{(0)}$ and the cooling rate $\gamma_{\rm c}^{(2)}$ as a function of the spontaneous cavity decay rate $\kappa$ for $\Omega=30 \, \Gamma$ and $\nu = \Gamma$.} \label{fig8}
\end{figure}

\subsubsection{Dependence on the laser Rabi frequency}

Fig.~\ref{fig4} shows the stationary state phonon number $m_{\rm ss}^{(0)}$ and the cooling rate $\gamma_{\rm c}^{(2)}$ as a function of the laser Rabi frequency $\Omega$. As suggested by Eq.~(\ref{mdot}), we find that there is no effective cooling, when the laser Rabi frequency $\Omega$ becomes very small. In the limit $\Omega \to 0$, the cooling rate $\gamma_{\rm c}^{(2)}$ tends for all three cooling resonances to zero. Although the stationary state phonon number $m_{\rm ss}^{(0)}$ might be relatively small, this case is of no interest, since the stationary state is reached only after a very long time. When $\Omega$ increases, also the cooling rate $\gamma_{\rm c}^{(2)}$ increases rapidly. Naively one might expect that increasing the laser Rabi frequency $\Omega$ further and further also increases the cooling rate further. This is not the case. As shown in Fig.~\ref{fig4}, the cooling process saturates relatively quickly and the stationary state phonon number remains more or less constant for very large $\Omega$.

When comparing all three cooling resonances, we see that the atom-cavity detuning $\delta_-$ yields the highest values of $m_{\rm ss}^{(0)}$ and is therefore of no practical interest. One reason for this can be found in Eqs.~(\ref{sideband}) and (\ref{heating}). For $\delta = \delta_-$, there is always a heating resonance relatively close by, which compensates some of the effects of the resonant cooling transition. Another reason for the relatively high values of $m_{\rm ss}^{(0)}$ for $\delta = \delta_-$ is that the applied laser field creates a relatively large population in the state $|\lambda_+ \rangle$ of the trapped particle, while the state $|\lambda_- \rangle$ remains less populated (cf.~Eq.~(\ref{zs})). As one can see from Eq.~(\ref{HI3}), for $\delta = \delta_-$, the resonant annihilation of a phonon and the creation of a cavity photon is accompanied by an atomic transition from the state $|\lambda_- \rangle$ into $|\lambda_+ \rangle$. When the average population in the state $|\lambda_- \rangle$ is relatively low, the atom is not well prepared to assist the cooling process when $\delta = \delta_-$. 

In contrast to this, the system is in general well detuned from all heating transitions, when the atom-cavity detuning equals either $\delta_0$ or $\delta_+$. Moreover, for $\delta = \delta_+$ and for $\delta = \delta_0$, resonant cooling transitions are accompanied by a $|\lambda_+ \rangle \to |\lambda_- \rangle$ and by a $|0 \rangle \to |1 \rangle$ or a $|1 \rangle \to |0 \rangle$ transition, respectively. Since the average population in the state $|\lambda_+ \rangle$ and in the atomic states $|0 \rangle$ and $|1 \rangle$, respectively, is relatively large (see again Eq.~(\ref{zs})), the laser driving prepares the trapped particle well to facilitate the annihilation of a phonon and to assist the cooling process when $\delta = \delta_+$ or $\delta = \delta_0$. Indeed, Fig.~\ref{fig4} shows that the atom-cavity detuning $\delta_+$ yields the lowest stationary state photon number $m_{\rm ss}^{(0)}$ for a relatively wide range of laser Rabi frequencies $\Omega$. For the concrete parameters in Fig.~\ref{fig4}, this applies when $\Omega$ lies roughly between 2 and $7 \, \Gamma$. For larger values of $\Omega$, we obtain the lowest stationary state phonon number when choosing $\delta = \delta_0$ (sideband cooling case). 

\subsubsection{Dependence on the phonon frequency}

Let us now have a closer look at the dependence of the cooling process on the phonon frequency $\nu$. To do so, we consider a relatively small and a relatively large value of $\Omega$, while keeping all other system parameters comparable to previous experimental parameters. As suggested by Fig.~\ref{fig4}, we choose $\Omega = 3 \, \Gamma$ (cf.~Fig.~\ref{fig5}) and $\Omega = 30 \, \Gamma$ (cf.~Fig.~\ref{fig6}). In Fig.~\ref{fig6}, we can easily identify two phonon frequencies $\nu$ for which certain cooling resonances (eg.~$\delta_-$) becomes identical to one of the heating resonances in Eq.~(\ref{heating}). When this applies, the cooling rate $\gamma_{\rm c}^{(2)}$ becomes very small (in some cases it even becomes negative which implies heating) and $m_{\rm ss}^{(0)}$ tends to infinity. Moreover, in both figures, the atom-cavity detuning $\delta_-$ yields the highest stationary state phonon numbers and is therefore of less practical interest than $\delta_0$ and $\delta_+$. For relatively small phonon frequencies $\nu$, the lowest stationary state phonon number is achieved, when the atom-cavity detuning equals $\delta_+$. For very strongly confined particles, it is better to choose $\delta = \delta_0$ (sideband cooling case). As one would expect, we notice that higher phonon frequencies allow to cool the trapped particle to significantly lower temperatures. 

\subsubsection{Dependence on the spontaneous cavity decay rate}

Finally, we discuss the dependence of $m_{\rm ss}^{(0)}$ and $\gamma_{\rm c}^{(2)}$ on the spontaneous cavity decay rate $\kappa$. As in the previous subsection, we choose $\Omega = 3 \, \Gamma$ (cf.~Fig.~\ref{fig7}) and $\Omega = 30 \, \Gamma$ (cf.~Fig.~\ref{fig8}). For a relatively wide range of experimental parameters, we find that the detuning $\delta_+$ yields the lowest stationary state phonon number (cf.~Figs.~\ref{fig7} and \ref{fig8}). This is especially then the case, when the spontaneous cavity decay rate $\kappa$ is relatively large. Although this is not illustrated here, we would like to add that the cooling transitions become over-damped when $\kappa$ becomes too large. In this case, the cooling becomes very inefficient and the stationary state phonon number $m_{\rm ss}^{(0)}$ increases rapidly. 

\section{Conclusions} \label{conclusions}

In this paper, we analyse cavity-mediated laser cooling for an atomic particle with external confinement in the direction of the cavity axis (cf.~Fig.~\ref{setup}). The Hamiltonian $H_{\rm I}$ of this system contains an atom-phonon-photon interaction term which gives rise to three sharp resonances with a minimum stationary state phonon number. For a wide range of experimental parameters, for example, when the spontaneous cavity decay rate $\kappa$ is relatively large or when the phonon frequency $\nu$ is relatively small, one should choose the atom-cavity detuning $\delta$ equal to $\delta_+$ in Eq.~(\ref{sideband}) in order to minimise the stationary state phonon number $m_{\rm ss}$ (cf.~Figs.~\ref{fig4}--\ref{fig8}). This resonance depends on the laser Rabi frequency $\Omega$ and is different from the usually considered resonance $\delta_0$ for laser-sideband cooling. 

To obtain an effective cooling rate $\gamma_{\rm c}$ and an analytical expression for the stationary state phonon number $m_{\rm ss}$ for the experimental setup which we consider in this paper (cf.~Eq.~(\ref{mdot})), we proceed as in Refs.~\cite{Tony,Norah,Tony3}. Starting from the standard quantum optical master equation, we derive linear differential equations -- so-called rate or cooling equations -- for the time evolution of expectation values. When taking a large enough number of expectation values into account, we obtain a closed set of  equations, which can be used to analyse the time evolution of the mean phonon number $m$ on a time scale given by $\eta^2$. The only assumption made in our calculations is that the atom-cavity coupling constant $g$ multiplied with the Lamb-Dicke $\eta$ is much smaller than at least one other experimental parameters (cf.~Eq.~(\ref{condi})). The condition in Eq.~(\ref{condi}) guarantees that the mean phonon number $m$ evolves on a much slower time scale than all the other relevant expectation values and allows us to obtain Eq.~(\ref{mdot}) via an adiabatic elimination.

Achieving very low stationary state phonon numbers for a single trapped particle requires a relatively large phonon frequency $\nu$, while very large spontaneous decay rates $\kappa $ and $\Gamma $ need to be avoided. Achieving relatively large cooling rates moreover requires a relatively large atom-cavity coupling constant $g$, since $\gamma_{\rm c}$ is proportional to $(\eta g)^2/\Gamma$. To overcome this problem, it might be interesting to study the cooling process of the experimental setup in Fig.~\ref{setup} when it contains many trapped particles \cite{Future}. Using the same arguments as in Section \ref{sec3} and diagonalising the system Hamiltonian with respect to its free energy and laser terms, one can show that many non-interacting particles experience exactly the same Mollow triplet of heating and cooling resonances as a single trapped particle. \\[1cm]
{\em Acknowledgement.} This work was supported by the UK Engineering and Physical Sciences Research Council EPSRC. Moreover, AB would like to thank P. Grangier for many inspiring discussions.


\begin{references}
\bibitem{sign}
It can be removed from the Hamiltonian, for example, by adding a minus sign to the excited electronic state $|1 \rangle$.
\end{references}
\end{document}